\documentclass[11pt,twoside]{article}
\usepackage{asp2010}

\newcommand{\Msun}{\mbox{$M_{\odot}$}}
\newcommand{\msun}{\mbox{$M_{\odot}$}}

\resetcounters

\begin{document}

\title{Constraining Mass-Loss \& Lifetimes of Low Mass, Low Metallicity AGB Stars}
\author{Philip Rosenfield$^1$, Paola Marigo$^1$, L\'eo Girardi$^2$
Julianne J.\ Dalcanton$^3$,
Alessandro Bressan$^4$,
Marco Gullieuszik$^2$,
Daniel Weisz $^{3,5}$,
Benjamin F.\ Williams$^3$,
Andrew Dolphin$^6$, and
Bernhard Aringer$^7$}

\affil{$^1$Dept. of Physics and Astronomy G. Galilei, University of Padova, Vicolo dell'Osservatorio 3, I-35122 Padova, Italy}
\affil{$^2$Osservatorio Astronomico di Padova -- INAF, Vicolo dell'Osservatorio 5, I-35122 Padova, Italy}
\affil{$^3$Dept. of Astronomy, University of Washington, Box 351580, Seattle, WA 98195, USA}
\affil{$^4$Astrophysics Sector, SISSA, Via Bonomea 265, I-34136 Trieste, Italy}
\affil{$^5$Hubble Fellow}
\affil{$^6$Raytheon Company, 1151 East Hermans Road, Tucson, AZ 85756, USA}
\affil{$^7$University of Vienna, Dept. of Astrophysics, Turkenschanzstra\ss e 17, A-1180 Wien, Austria}

\begin{abstract}
The evolution and lifetimes of thermally pulsating asymptotic giant branch (TP-AGB) stars suffer from significant uncertainties. We present a detailed framework for constraining model luminosity functions of TP-AGB stars using resolved stellar populations. We show an example of this method that compares various TP-AGB mass-loss prescriptions that differ in their treatments of mass loss before the onset of dust-driven winds (pre-dust). We find that models with more efficient pre-dust driven mass loss produce results consistent with observations, as opposed to more canonical mass-loss models. Efficient pre-dust driven mass-loss predicts for [Fe/H] $\lesssim -1.2$, lower mass TP-AGB stars (M $\lesssim 1 \Msun$) must have lifetimes less than about 1.2 Myr.
\end{abstract}

\section{Introduction}
Resolved stellar populations (RSPs) are powerful laboratories for understanding uncertain phases of stellar evolution, including the thermally pulsating asymptotic giant branch \citep[TP-AGB; e.g.,][]{Marigo2008}. Understanding TP-AGB stars in dwarf galaxies shed light on the processes of galaxies at larger redshift where resolving the stellar content is impossible. In addition, TP-AGB stars can also contribute $\sim20\%$ of a galaxy's NIR light \citep{Melbourne2012}. 

Despite their importance, TP-AGB models uncertainly predict the brightness distribution (i.e., the luminosity function; LF) of AGB stars beyond the MCs \citep[e.g., see discussion in][]{Girardi2010}. The differences between observed and modeled LFs of TP-AGB stars found in a RSP is mainly due to model TP-AGB lifetimes, which are primarily set by mass loss \citep[e.g.,][]{Vassiliadis1993}.

Recently, the Hubble Space Telescope has optically imaged $\sim70$ nearby galaxies that together house thousands of TP-AGB stars \citep{Dalcanton2009}. From this sample, 23 galaxies also have HST/NIR imaging \citep{Dalcanton2012agbsnap}. These datasets provide over 1000, low metallicity, TP-AGB stars which we use to constrain model TP-AGB mass-loss prescriptions, and thus TP-AGB model lifetimes. Here, we outline a method expanded from \citet{Girardi2010} to constrain uncertain parameters of TP-AGB models with observations of RSPs, and illustrate the method with summary analysis presented in \citet{Rosenfield2014}. By ensuring a large model parameter search space, this method is a step toward the goal of a fully Bayesian framework for constraining stellar evolution models.

\section{Method}
\begin{figure}
\includegraphics[width=3in]{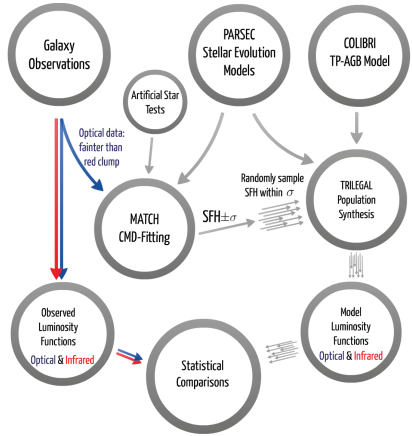}
\caption{Schematic diagram of the method to statistically compare the predicted TP-AGB model luminosity function to that observed (see text).}
\label{fig1}
\end{figure}

Fig.\ \ref{fig1} is a schematic diagram of the method to statistically compare TP-AGB models with observations. We are primarily interested the shape of the LF brighter than the tip of the red giant branch (TRGB). However, this method is applicable to higher dimensional data, such as color-magnitude diagrams (CMDs) and color-color diagrams of any RSP. Briefly, we compare the observed LFs to synthetic RSPs that are created based on the most likely star formation history (SFH) of the data and a specified TP-AGB mass-loss prescription.

\subsection{Observed Luminosity Functions}
Beginning in the top left of Fig.\ \ref{fig1}, luminosity functions are obtained from the optical and NIR data. Using the optical data, we also calculate at least 100,000 artificial star tests for each image \citep[described in e.g.,][]{Dalcanton2009}, which are used to establish a faint magnitude limit (set at 90\% completeness) to compare LFs, are applied to the stellar evolution models when deriving the SFH, and applied to the model LFs to correct for completeness. In addition, RGB stars are counted in order to scale the population synthesis models that provide the model LFs  \citep[RGB stars are usually identified by a CMD box, c.f.,][]{Girardi2010, Rosenfield2014, Melbourne2012}.

\subsection{Model Luminosity Functions}

\subsubsection{Stellar Evolution Models}
Two stellar evolution codes are used in recovering the SFH of the galaxy and to create synthetic stellar populations (top and right of Fig.\ \ref{fig1}).

\noindent \emph{Padova and Trieste Stellar Evolution Code}:
PARSEC v1.1 \citep[][]{Bressan2012} is an updated Padova Stellar Evolution code, containing models spanning $Z=0.0001-0.06$, $M=0.1-12\msun$, from the Pre-MS to either the TP-AGB or core Carbon ignition.

\noindent \emph{COLIBRI}:
Following the first thermal pulse on the AGB, COLIBRI takes over the stellar evolution calculations from PARSEC \citep[][]{Marigo2013}. Briefly, COLIBRI optimizes the ratio
of physical accuracy over computational issues typical of TP-AGB models, counting on a detailed envelope model in which the molecular chemistry of $>$800 species and gas opacities are computed on-the-fly \citep{Marigo2009}.

\subsubsection{Star Formation Histories}
\label{sec_sfh}

SFHs are derived using the CMD-fitting MATCH package \citep{Dolphin2000} using the deeper optical data. MATCH finds the most-likely CMD that fits the optical CMD based on a given IMF, binary fraction, artificial star tests, and stellar models. In this work, we adopt a \citet{Kroupa2001} IMF and a binary fraction of 0.35.

The input stellar models used in MATCH are from PARSEC, that is, we exclude TP-AGB models in the SFH derivation and give no weight to the regions of the observed CMD above the measured TRGB and redder than the main sequence.

Uncertainties in the SFHs propagate to a spread in the predicted LF (see Fig.\ \ref{fig2}). We use the most-likely SFH and its random uncertainties as inputs to the population synthesis models \citep[for complete details see][]{Dolphin2013}.

\subsubsection{Population Synthesis}
\label{sec_popsynth}
We model the photometry of RSPs with the TRILEGAL population synthesis code \citep{Girardi2005}. TRILEGAL takes as input the PARSEC and COLIBRI stellar evolution models, an IMF, binary fraction, and the time evolution of metallicity and star formation rate. Importantly, TRILEGAL also simulates the $L-T_{\rm eff}$ variations due to the thermal pulse cycles on the TP-AGB \citep{Girardi2007}. The TRILEGAL input parameters are set to remain consistent with the parameters listed in Sec.\ \ref{sec_sfh}. Finally, to fully explore the possible parameter space, we randomly draw SFHs from the most-likely SFH and its given uncertainties.

Each synthetic RSP is scaled to match the number of RGB stars identified from the observations. The scaled luminosity functions are finally corrected for completeness based on the artificial star tests.

\subsection{Statistical Comparisons}
The resulting set of model and observed LFs are compared by using a statistic similar to $\chi^2$ but for a Poisson probability distribution \citep[e.g.,][]{Dolphin2002}.

\section{Comparing Three Mass Loss Prescriptions}
\label{sec_mloss}
\begin{figure}
\includegraphics[width=4.5in]{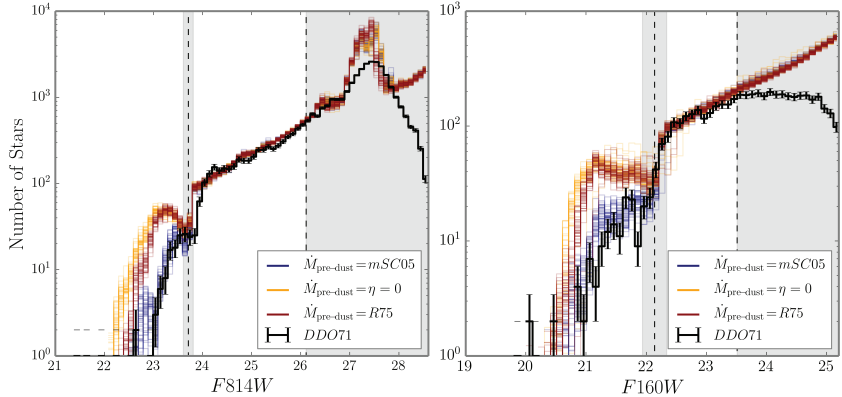}
\caption{LFs of DDO71 with three mass loss prescriptions $\eta=0$, $R75$, and $mSC05$. The spread in model LFs is due to sampling the most-likely SFH and its uncertainties. The faintest shaded region marks the 90\% completeness limit, and brighter shaded region denotes the TRGB \citep[][]{Dalcanton2009, Dalcanton2012agbsnap}.}
\label{fig2}
\end{figure}

\begin{figure}
\includegraphics[width=2.5in]{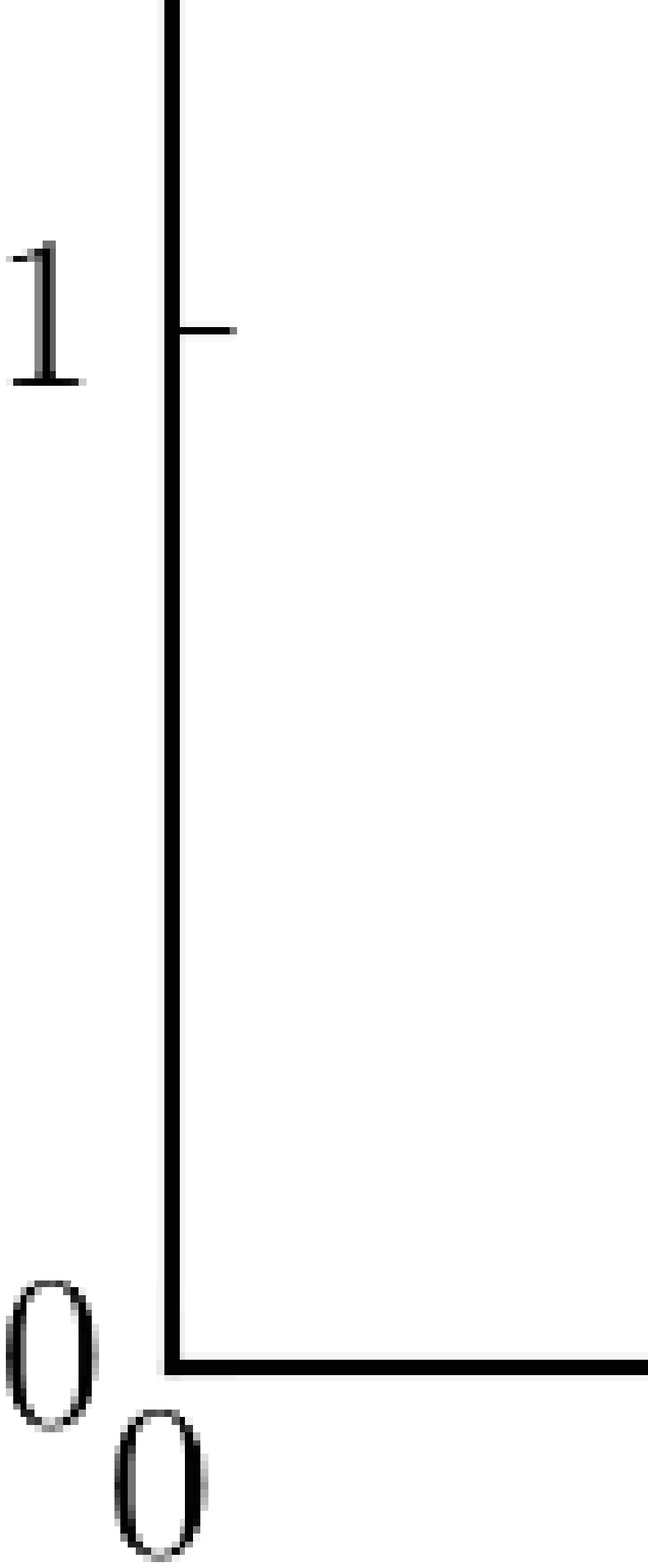}
\includegraphics[width=2.5in]{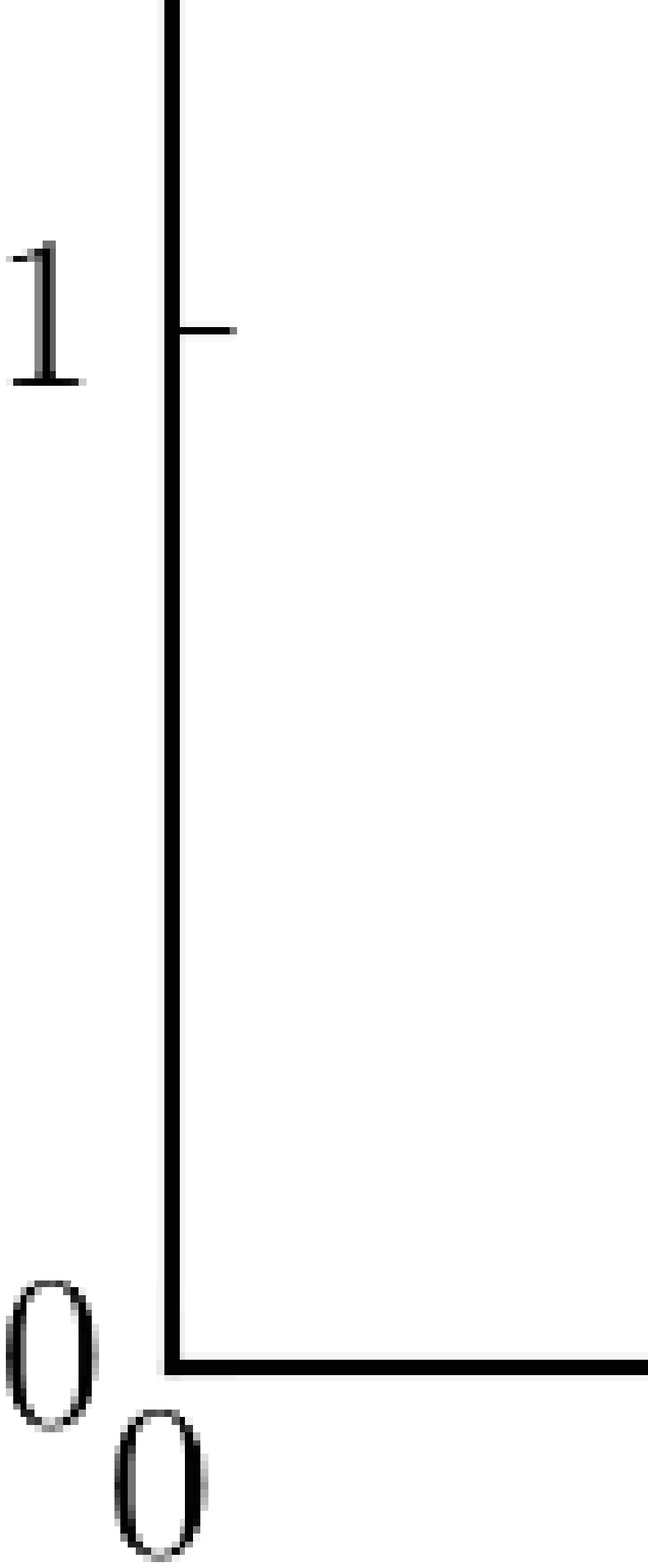}
\caption{Lifetime of TP-AGB stars as a function of initial mass for several metallicities predicted by the $mSC05$ model. Left: the total lifetime of the pre-dust phase. Right: the lifetime of the TP-AGB phase brighter than the TRGB (measurable from observations).}
\label{fig3}
\end{figure}

As an example of the success of this method we summarize recent findings from \citet{Rosenfield2014} which emphasized the importance of mass loss on the AGB before the onset of dust-driven winds. In this work, we compared the results of this method using three pre-dust mass loss regimes: 1) no pre-dust mass loss ($\eta=0$); 2) The canonical \citep[][R75]{Reimers1975} mass-loss with efficiency $\eta=0.4$; and 3) a modified version of \citep{Schroeder2005} based on \citep[][$mSC05$]{Cranmer2011}. Fig.\ \ref{fig2} shows the model LFs with each mass loss prescription as a visual guide. Clearly, ignoring pre-dust mass loss or simply using $R75$ over-predicts the number of TP-AGB stars, and $mSC05$ is a promising fit. In terms of the Poisson likelihood parameter, $mSC05$ is a factor of 2-3 lower than the other prescriptions. Using $mSC05$, we show the resulting lifetimes of TP-AGB stars as a function of mass for a range of metallicities in Fig.\ \ref{fig3}.

\section{Conclusions}
We have presented a framework for constraining stellar evolution models and showed its use in constraining low mass, low metallicity, TP-AGB mass-loss rates, which emphasizes the importance of including efficient pre-dust mass loss in TP-AGB models. Specifically, $mSC05$ predicts for [Fe/H] $\lesssim -1.2$, lower mass TP-AGB stars (M $\lesssim 1 \Msun$) must have lifetimes less than about 1.2 Myr.

\acknowledgements
Based on observations made with the NASA/ESA Hubble Space Telescope, obtained from the Data Archive at the Space Telescope Science Institute (STScI), which is operated by the Association of Universities for Research in Astronomy, Inc., under NASA contract NAS 5-26555. We acknowledge support from Progetto di Ateneo 2012 (University of Padova, ID: CPDA125588/12), HST GO-10915, and  the ERC Consolidator Grant (ID: 610654, project STARKEY). Support for DRW is from NASA through Hubble Fellowship grants HST-HF-51331.01 awarded by the STScI.

\bibliography{Rosenfield}

\end{document}